\documentclass[aps,prb,preprint,nopacs,superscriptaddress,]{revtex4}

\usepackage{graphicx}
\usepackage{verbatim}
\usepackage{mathrsfs}
\usepackage{gensymb}
\usepackage{color}
\usepackage{ulem}
\usepackage{times}

\usepackage{etoolbox}
\pretocmd{\abstractname}{\newpage}{}{}

\usepackage{amsmath,amsfonts,amssymb}
\usepackage{graphicx}

\newcommand{\bee}{\begin{equation}}
\newcommand{\ee}{\end{equation}}

\def\3{2.5in}    
\def\2{2.5in}
\def\4{3.0in}\def \beq {\begin{equation}}
\def \eeq {\end{equation}}
\begin{document}

\title{Helicoid-arc van Hove singularities in topological chiral crystals}

\author{Daniel S. Sanchez $^*$\footnote[0]{*These authors contributed equally to this work.}}
\affiliation {Laboratory for Topological Quantum Matter and Advanced Spectroscopy (B7), Department of Physics, Princeton University, Princeton, New Jersey 08544, USA}\affiliation{Center for Quantum Devices and Microsoft Quantum-Copenhagen, Niels Bohr Institute, University of Copenhagen, Universitetsparken 5, Copenhagen 2100, Denmark}

\author{Tyler A. Cochran $^*$}
\affiliation {Laboratory for Topological Quantum Matter and Advanced Spectroscopy (B7), Department of Physics, Princeton University, Princeton, New Jersey 08544, USA}

\author{Ilya Belopolski} 
\affiliation {Laboratory for Topological Quantum Matter and Advanced Spectroscopy (B7), Department of Physics, Princeton University, Princeton, New Jersey 08544, USA}

\author{Zi-Jia Cheng} 
\affiliation {Laboratory for Topological Quantum Matter and Advanced Spectroscopy (B7), Department of Physics, Princeton University, Princeton, New Jersey 08544, USA}

\author{Xian P. Yang} 
\affiliation {Laboratory for Topological Quantum Matter and Advanced Spectroscopy (B7), Department of Physics, Princeton University, Princeton, New Jersey 08544, USA}

\author{Yiyuan Liu}
\affiliation {International Center for Quantum Materials, School of Physics, Peking University, Beijing, China}

\author{Xitong Xu}
\affiliation{International Center for Quantum Materials, School of Physics, Peking University, China}

\author{Kaustuv Manna}
\affiliation {Max Planck Institute for Chemical Physics of Solids, Dresden, Germany}\affiliation{Department of Physics, Indian Institute of Technology Delhi,
Hauz Khas, New Delhi 110016, India}

\author{Jia-Xin Yin}
\affiliation {Laboratory for Topological Quantum Matter and Advanced Spectroscopy (B7), Department of Physics, Princeton University, Princeton, New Jersey 08544, USA}

\author{Horst Borrmann}
\affiliation {Max Planck Institute for Chemical Physics of Solids, Dresden, Germany}

\author{Alla Chikina}
\affiliation {Swiss Light Source, Paul Scherrer Institute, Villigen, Switzerland}

\author{Jonathan Denlinger}
\affiliation {Advanced Light Source, Lawrence Berkeley National Laboratory, Berkeley, CA, 94720, USA}

\author{Vladimir N. Strocov}
\affiliation {Swiss Light Source, Paul Scherrer Institute, Villigen, Switzerland}

\author{Claudia Felser}
\affiliation{Max Planck Institute for Chemical Physics of Solids, Dresden, Germany}

\author{Shuang Jia}
\affiliation{International Center for Quantum Materials, School of Physics, Peking University, China}

\author{Guoqing Chang $^\dagger$}
\affiliation {Division of Physics and Applied Physics, School of Physical and Mathematical Sciences, Nanyang Technological University, Singapore}

\author{M. Zahid Hasan $^\dagger$\footnote[0]{$^\dagger$ To whom correpsondence should be addressed: mzhasan@princeton.edu, guoqing.chang@ntu.edu.sg}}
\affiliation{Laboratory for Topological Quantum Matter and Advanced Spectroscopy (B7), Department of Physics, Princeton University, Princeton, New Jersey 08544, USA}\affiliation{Princeton Institute for Science and Technology of Materials, Princeton University, Princeton, New Jersey 08544, USA}\affiliation{Lawrence Berkeley National Laboratory, Berkeley, CA 94720, USA}

\maketitle

\textbf{Van Hove singularity are electronic instabilities that lead to many fascinating interactions, such as superconductivity and charge-density waves. And despite much interest, the nexus of emergent correlation effects from van Hove singularities and topological states of matter remains little explored in experiments. By utilizing synchrotron-based angle-resolved photoemission spectroscopy and Density Functional Theory, here we provide the first discovery of the helicoid quantum nature of topological Fermi arcs inducing van Hove singularities. In particular, in topological chiral conductors RhSi and CoSi we directly observed multiple types of inter- and intra-helicoid-arc mediated singularities, which includes the type-I and type-II van Hove singularity. We further demonstrate that the energy of the helicoid-arc singularities are easily tuned by chemical engineering. Taken together, our work provides a promising route to engineering new electronic instabilities in topological quantum materials.}\\


Emergent correlated quantum states of matter arise in a large variety of platforms \cite{kmoore, revHK, revQZ, rev6, rev6a, rev7, rev5}. Cuprate and iron-based compounds host superconductivity at high temperatures \cite{Cupprates_VHS}, while graphene superlattice structures exhibit correlated insulating, superconducting, and magnetic states \cite{TBG_VHS}. Furthermore, time-reversal breaking charge density waves exist in many  kagome superconductors \cite{rev5, Kagome_SC}. In all such instances, the unconventional phenomenology is attributed to the large density of states (DOS) at (or near) the Fermi level, which becomes unstable due to the emergence of electronic order. In twisted bilayer graphene, one contribution to the large DOS arises from the flat bands at specific magic twist angles \cite{TBG_VHS}. Another contribution, also pertinent to the cuprate and kagome materials, are the van Hove singularities in the bare electronic structure \cite {Graphene_VHS, Graphene_NanoARPES}. A recent theoretical study proposed that van Hove singularities may arise in two-types \cite{Magic_VHS, TypeII_VHS}. The conventional (type-I) van Hove singularity has a saddle point in the electronic band structure that is located at time-reversal invariant momenta (TRIM). The projection onto the surface Brillouin zone (surface BZ) yields type-I van Hove singularities located at the surface high-symmetry points. Unconventional (Type-II) van Hove singularities occur at arbitrary momenta in the surface BZ, which corresponds to a saddle point dispersion that is no longer constrained to reside at a TRIM. Type-II van Hove singularities are predicted to be an essential ingredient for realizing topological odd-parity superconductivity \cite{TypeII_VHS, ChiralTSC_Correlations}. The search for quantum materials with type-II van Hove singularities remains an ongoing research direction.

At the same time, topological materials have gained great interest for their own distinct electronic properties, such as protected surface states and giant/quantized transport and optical susceptibility \cite{kmoore, revHK, revQZ, rev6, rev6a, rev7, rev5}. Specifically, Weyl semimetals exhibit gapless Weyl fermions in their bulk electronic structure, that carry a topologically protected chiral charge of Berry curvature flux \cite{rev6, rev6a, rev5}. As a result, Fermi arc states appear as non-closed Fermi pockets that connect the projections of Weyl fermions on the surface BZ \cite{TaAs1, TaAs2, ARPES-TaAs1, ARPES-TaAs2}. In this class of materials, the topology can give rise to a large anomalous Hall effect and, under an external magnetic field, the chiral anomaly \cite{rev6, rev6a, rev5, chiralmagenticeffect1, chiral_photogalvanic}. What is more, the Weyl fermion dispersion in structurally chiral crystals is generalized to a chiral fermion dispersion that is pinned by time-reversal symmetry at high-symmetry points in the bulk Brillouin zone (bulk BZ) and may possess a two-fold or higher-fold degeneracy with a chiral charge $|C|>1$ \cite{unconventionalWeyl, filling_constraint1, Ben1, KramersWeyl2}. Such quantum materials are known as topological chiral crystals ( or topological chiral conductors) and host additional novel electronic properties, such helicoid-arc quantum states and unconventional photocurrents \cite{HelicodalFermiArcs, Chang_QPGE}.

In this work, we merge the concepts of van Hove singularities and topological conductors, showing multiple types of helicoid-arc singularities on the surface of RhSi and CoSi crystals \cite{RhSi, CoSi, CoSi_Nature1,CoSi_Nature2,CoSi_Transport}. Furthermore, we theoretically show that the intrinsic helicoid nature of Fermi arcs \cite{HelicodalFermiArcs} guarantees the existence of van Hove singularities at high-symmetry points in the BZ. Using synchotron-based angle-resolved photoemission spectroscopy \cite{SLS_SXARPES}, we probe the helicoid-arc dispersion and identify the van Hove singularities both above and below the Fermi level in topological chiral conductors RhSi and CoSi, respectively. Finally, with our material specific chemical gating, we show that it is possible to tune the Fermi level to the van Hove energy.



\begin{figure}
\centering
\includegraphics[width=135mm]{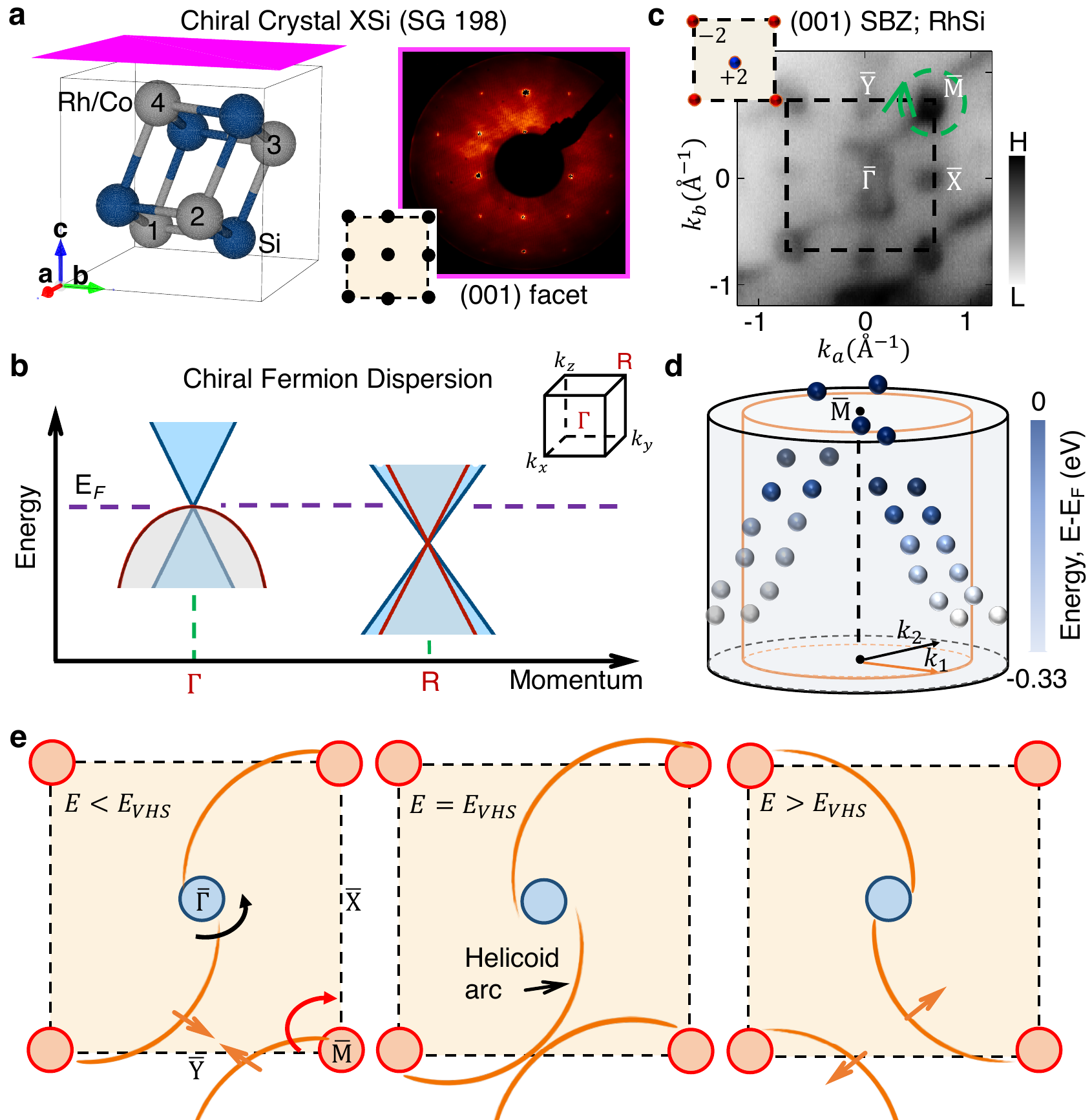}
\caption{\textbf{Helicoid-arcs in topological chiral conductors and van Hove physics} 
\textbf{a}, Chiral crystals RhSi and CoSi crystallize in space group 198. Left Panel: The real space arrangement of Rh- (Co-) atoms in the unit cell exhibit a chiral arrangement (illustrated by numbers: $1-4$). The above magenta plane corresponds to the (001) surface. Right panel: Low-energy electron diffraction (LEED) pattern of the (001) surface. Bottom inset: Illustration of the expected LEED pattern for the (001) surface. \textbf{b}, Schematic of the three-fold and four-fold chiral fermions at the $\Gamma$ and $R$ high-symmetry points, respectively. Top inset: Illustration of the bulk BZ. \textbf{c}, VUV-ARPES measured Fermi surface of RhSi. The black-dashed line marks boundary of the surface BZ (SBZ), with high-symmetry points labeled. \textbf{d}, 3D energy-dispersion data plot along two concentric loops enclosing the $\bar{M}$-point. A clockwise spiraling around the $\bar{M}$-axis is observed as the binding energy is tuned towards the Fermi level. \textbf{e}, Cartoon illustration showing a type-I van Hove singularity generated via an inter-helicoid-arc touching (middle panel). As a function of binding energy, the helicoid-arcs approach each other (left panel), touch at the van Hove energy (middle panel: $E_{VHS}$), and move away with a different connectivity (right panel).
}
\label{Fig1} 
\end{figure}

RhSi and CoSi are non-magnetic iso-structural single crystals that grow in a simple cubic Bravais lattice in non-symmorphic space group $P2_{1}13$ (No. 198). Per unit cell, the Rh (Co)- and Si- atoms are arranged along the crystal $c$-axis with a right- or left-handed chiral pattern (left panel: Fig.~\ref{Fig1}\textbf{a}). To obtain high-resolution measurements of the electronic band structure with ARPES, an \textit{in situ} surface preparation procedure using cycles of Argon sputtering and high-temperature annealing was implemented (see Methods). To ensure the sample preparation procedure was not destructive to the surface crystalline order, low-energy electron diffraction (LEED) measurements were performed. The observed LEED spots are arranged in a square-like pattern for electron energies $87-150\,keV$, consistent with the symmetries expected for the (001) surface termination of RhSi and CoSi (right panel: Fig.~\ref{Fig1}\textbf{a}). And RhSi and CoSi possess a similar electronic band-structure, as shown by previous \textit{ab initio} calculations and experiments. In the spinless limit, a de facto $C=+2$ and $C=-2$ chiral charge is carried by the three-fold chiral fermion at $\Gamma$ and four-fold chiral fermion at $R$, respectively, Fig.~\ref{Fig1}\textbf{b}. Now, the projection of the bulk chiral charges onto the (001) surface results in a time-reversed pair of long Fermi arc surface states extending diagonally across the surface BZ ($\bar{M}-\bar{\Gamma}-\bar{M}$), consistent with our ARPES measurements, Fig.~\ref{Fig1}\textbf{c}. In particular, since the $\bar{\Gamma}$ and $]\bar{M}$ pockets enclose a projected chiral charge of $C=+2$ and $C=-2$, respectively, two Fermi arcs are expected to terminate at each pocket. By plotting the evolution of the long Fermi arcs as a function of binding energy along a closed loop at the $\bar{M}$-point, its helicoid property is directly visualized \cite{arcDetect1}. Additionally, it is resolved in experiment that the rotation direction at the termination points of the helicoid-arc is determined by the chirality of the enclosed projected chiral charge, Fig.~\ref{Fig1}\textbf{d}. The counter-rotating ends of the helicoid-arc results in a non-trivial winding in the surface BZ as the binding energy is varied.  To-that-end, the helicoid-arcs are allowed to touch at various points in surface BZ, potentially generating various types of van Hove singularities. At time-reversal invariant momenta, van Hove singularities, classified as type-I, are created through an inter-helicoid-arc interaction, Fig.~\ref{Fig1}\textbf{e}. A suspected novel result accompanying such a van Hove singularity is a change to the helicoid-arc connectivity as the binding energy is tuned across the van Hove energy ($E_{VHS}$). To realize in experiment van Hove singularities generated by helicoid-arcs, we begin by further applying the ARPES technique to RhSi and chemically-doped RhSi.

\begin{figure}
\centering
\includegraphics[width=165mm]{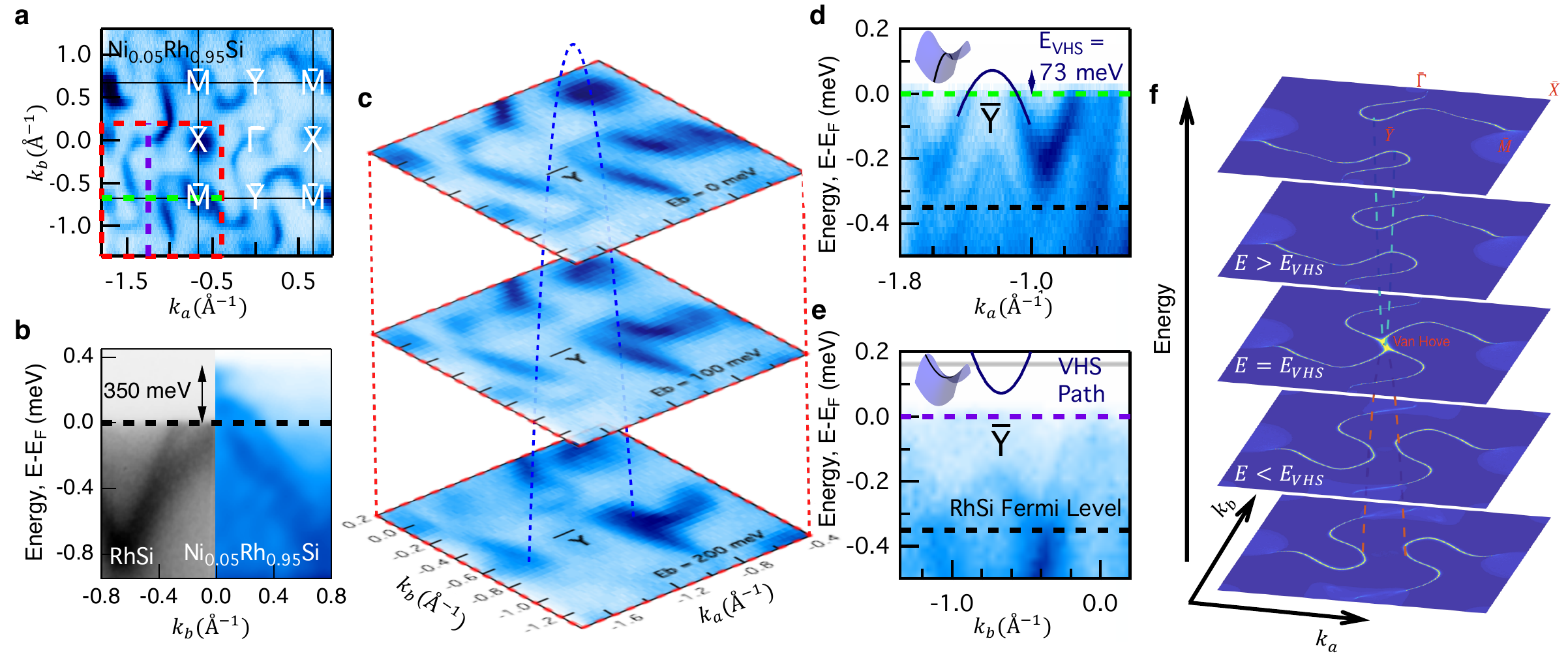}
\caption{\textbf{Type-I helicoid-arc van Hove singularity in topological chiral conductor RhSi.}
\textbf{a}, VUV-ARPES measured Fermi surface of Ni$_{0.05}$Rh$_{0.95}$Si with an incident photon energy of $85\,eV$. The surface Brillouin zone boundary is marked by black-line grid, with the high-symmetry points labeled. \textbf{b}, Energy-dispersion cut along the $\bar{\Gamma}$-pocket for un-doped RhSi (left side) and electron-doped RhSi (Ni$_{0.05}$Rh$_{0.95}$Si, right side). Relative to RhSi, the Fermi level in Ni$_{0.05}$Rh$_{0.95}$Si is shifted upwards by approximately $0.35\,eV$. \textbf{c}, Constant-energy contours for Ni$_{0.05}$Rh$_{0.95}$Si, region illustrated by the red-dashed box in panel (a). The helicoid-arcs near the $\bar{M}-\bar{Y}-\bar{M}$ high-symmetry line approach each other as the binding energy is varied, and are on track to touch at the $\bar{Y}$-point. \textbf{d}, \textbf{e}, Energy dispersion cuts along the $\bar{Y}$-point. Two orthogonal directions are studied and fitted with a parabolic function to further illustrate the saddle-point dispersion of the helicoid-arcs and find the van Hove energy, which is approximately $0.073\,eV$ above Fermi level of Ni$_{0.05}$Rh$_{0.95}$Si. The black dashed line marks the Fermi level of RhSi. \textbf{f}, \textit{Ab initio} calculated constant-energy contours visualizing a type-I helicoid-arc van Hove singularity at the $\bar{Y}$-point. The helicoid-arc connectivity changes as the binding energy is tuned through the van Hove energy ($E_{VHS}$). 
}
\label{Fig2}
\end{figure}

Single crystals of chemically doped RhSi were grown with the composition Ni$_{x}$Rh$_{1-x}$Si, where $x$ is varied by varying the Ni-concentration. The result of higher Ni concentrations is an electron doping that raises the Fermi level in RhSi. With a further optimized \textit{in situ} surface preparation procedure, higher-quality ARPES measurements were obtained. And using incident photons with an energy of $85\,eV$, Ni$_{0.05}$Rh$_{0.95}$Si was observed to possess an electronic band-structure that is consistent with that of RhSi, Fig.~\ref{Fig2}\textbf{a}. Particularly important for this study, strong signatures of time-reversed helicoid-arcs spanning diagonally across the $\bar{M}-\bar{\Gamma}-\bar{M}$ direction on the (001) surface BZ was observed. Moreover, the spiraling of the helicoid-arcs at the $\bar{\Gamma}$- and $\bar{M}$-points was resolved to be counter-clockwise and clockwise direction as the binding energy is increased, respectively. An energy-dispersion cut along the $\bar{\Gamma}$-pocket for un-doped and doped RhSi resolves signatures of the three-fold chiral fermion pinned at the $\Gamma$-point. Suggesting that the only observable change to the electronic band-structure of doped RhSi is that the Fermi level is raised by approximately $0.35\,eV$, Fig.~\ref{Fig2}\textbf{b}. ARPES resolved constant-energy contours near the Fermi level of doped RhSi tracks the momentum space trajectory of the helicoid-arcs by the $\bar{M}-\bar{Y}-\bar{M}$ line, Fig.~\ref{Fig2}\textbf{c}. Starting from low binding energies and then moving towards the Fermi level, the two helicoid-arcs are observed moving towards the $\bar{Y}$-point and each other. Guides to the eye tracking helicoid-arcs as a function of binding energy illustrates the possibility of a type-I van Hove singularity occurring slightly above the Fermi level at the $\bar{Y}$-point. Fitting the helicoid-arcs near the $\bar{Y}$-point with a parabolic function shows that the measurements on Ni$_{0.05}$Rh$_{0.95}$Si resolves the lower type-I van Hove branch, further suggesting that the van Hove energy is approximately $0.073\,eV$ above the Fermi level, Fig.~\ref{Fig2}\textbf{e}. Inspection of an energy-dispersion cut along the $\bar{Y}$-point in an orthogonal direction resolves no bands near the Fermi level, consistent with the upper branch of the type-I van Hove singularity residing above the Fermi level, Fig.~\ref{Fig2}\textbf{f}. \textit{Ab initio} calculated constant energy contours for RhSi are nominally consisted with ARPES measurements and independently indicates that a type-I van Hove singularity is generated by the helicoid-arcs at the $\bar{Y}$-point. The helicoid-arc connectivity is observed in calculation to change as the binding energy is tuned through the van Hove energy.

\begin{figure}[t]
\includegraphics[width=165mm]{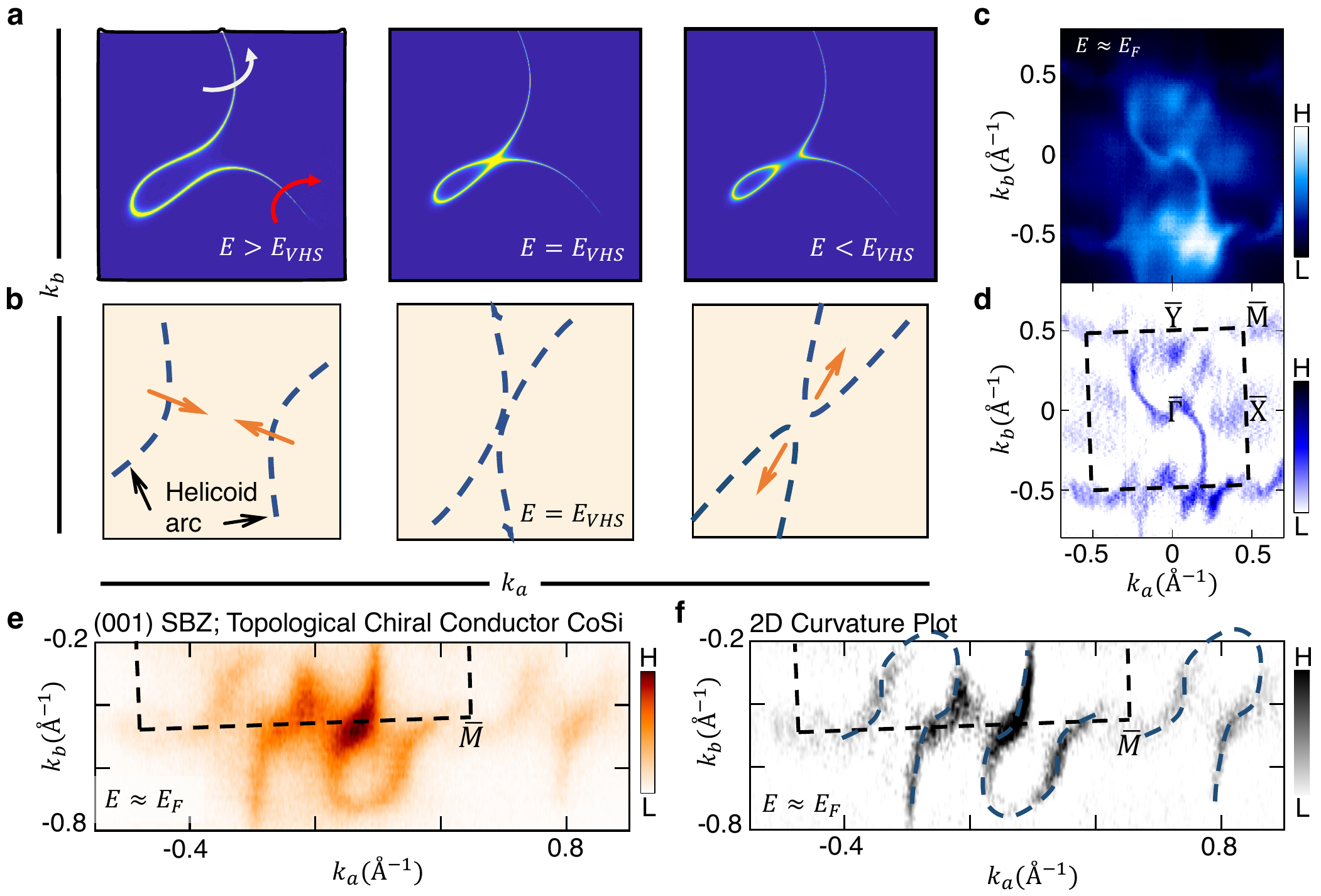}
\caption{ \textbf{Unconventional type-II helicoid-arc van Hove singularity in topological chiral conductor CoSi.} 
\textbf{a}, \textit{Ab initio} calculated constant-energy contours of the helicoid-arc surface state near the $\bar{M}$ point on the (001) surface Brillouin zone in CoSi. The calculation shows that as the binding energy is varied from high to low (right to left panel), at a generic location in the surface Brillouin zone, the helicoid-arc generates an unconventional type-II van Hove singularity at $E=E_{VHS}$ (middle panel). The counterclockwise and clockwise white and red arrows, respectively, further illustrates the helicoid-winding property of the topological surface state near their termination points. \textbf{b}, Cartoon illustration of the helicoid-arc (blue dashed-line) generating a type-II van Hove singularity predicted in calculation. \textbf{c}, VUV-ARPES measured constant energy contour near the Fermi level ($E_{F}$) of CoSi with an incident photon energy of 87eV. \textbf{d}, corresponding 2D curvature plot of panel (c). Black-dashed square illustrates the (001) surface Brillouin zone boundary, with the surface high-symmetry points labeled. \textbf{e}, High-resolution Fermi surface of the long-winding helicoid-arc surface states near the bottom surface Brillouin zone edge. \textbf{f}, Corresponding 2D curvature plot with blue-dashed lines as guides to the eye for the helicoid-arcs in CoSi.}
\label{Fig3}
\end{figure}

In CoSi, an unconventional type of van Hove singularity is realizable. Presented in Fig.~\ref{Fig3}\textbf{a} are the \textit{ab initio} calculated of constant-energy contours near the $\bar{\Gamma}-\bar{M}$ direction in the (001) surface BZ of CoSi. Observed is a helicoid-arc connecting the $\bar{\Gamma}$- and $\bar{M}$-point. And the counter-clockwise and clockwise rotation of the helicoid-arc at the $\bar{\Gamma}$- and $\bar{M}$-points, respectively, results in an intra-type-II van Hove singularity as the binding energy is decreased, Fig.~\ref{Fig3}\textbf{b}. VUV-ARPES measurements using incident photons with an energy of $87\,eV$ resolved the surface electronic states with high-contrast, Fig.~\ref{Fig3}\textbf{c},\textbf{d}. Similar to RhSi, the chiral charges $C=\pm{2}$ pinned at bulk BZ center and corners project onto the surface Brillouin zone center $\bar{\Gamma}$ and corner $\bar{M}$. As a result, time-reversed pairs of helicoid-arcs are measured extending diagonally from the surface zone center to the corners. What is more, the helicoid-arcs are observed to curve around, kink and extend into neighboring surface BZ. High-resolution measurements near the lower region of the surface Brillouin zone shows with high-contrast the trajectory and termination of the helicoid-arcs at the $\bar{M}$ points, Fig.~\ref{Fig3}\textbf{e}. A corresponding 2D curvature plot further emphasizes the topological states of interest, Fig.~\ref{Fig3}\textbf{f}.


Further study of the helcoid-arcs in CoSi as the binding energy is lowered shows remarkable agreement with our \textit{ab initio calculations}, Fig.~\ref{Fig4}\textbf{a}. It is observed that as the binding energy is lowered, points along the helicoid-arc contract, touch, and then pull away. What is more, the intra-helical-arc touching point occurs at a generic point in the surface BZ at a binding energy of approximately $-0.03\,eV$ below the Fermi level,  Fig.~\ref{Fig4}\textbf{b}, Fig.~\ref{Fig4}\textbf{c}. The curvature plots of energy-dispersion cuts that slice, in orthogonal directions, through the helical-arc touching point shows spectroscopic evidence of a saddle-point dispersion, Fig.~\ref{Fig4}\textbf{d}. APRES measurements presented here provide a direct visualization of type-II van Hove singularity that is mediated by an intra-helicoid-arc interaction in CoSi. Together, RhSi and CoSi provide a suitable material family for realizing type-I and type-II van Hove singularities that emerge due to inter- and intra-interaction of helicoid-arc states, Fig.~\ref{Fig4}\textbf{e}. And as was demonstrated by the Ni-doped RhSi results presented above, and by other studies on Fe-doped CoSi, the Fermi level can be experimentally varied by chemical doping. By calculation, it is predicted that lightly chemically doped Ni$_{0.07}$Rh$_{0.93}$Si (Fig.~\ref{Fig4}\textbf{f}) and Fe$_{0.01}$Co$_{0.99}$Si (Fig.~\ref{Fig4}\textbf{g}) will have their Fermi level shifted to the van Hove energy. Such low-levels of chemical doping is experimentally feasible.

\textbf{Acknowledgments:} Work at Princeton University and Princeton-led synchrotron based ARPES measurements were supported by the United States Department of Energy (US DOE) under the Basic Energy Sciences program (grant number DOE/BES DE-FG-02-05ER46200). This research used resources of the Advanced Light Source, which is a DOE Office of Science User Facility under contract No. DE-AC02-05CH11231. We acknowledge the Paul Scherrer Institut, Villigen, Switzerland for provision of synchrotron radiation beamtime at the ADRESS beamline of the Swiss Light Source. G.C. would like to acknowledge the support of the National Research Foundation, Singapore under its NRF Fellowship Award No. NRF-NRFF13-2021-0010 and the Nanyang Assistant Professorship grant from Nanyang Technological University. T. A. C. was supported by the National Science Foundation Graduate Research Fellowship Program under Grant No. DGE-1656466.

\textbf{Author contributions:} D. S. S., T. A. C., G. C. and M. Z. H. conceived the project. D. S. S., and T. A. C. conducted the ALS ARPES experiments with the close assistance from I. B. Z. C., X. P. Y., and J. D., and in consultation with M. Z. H.; T. A. C. and I. B. conducted the SLS ARPES experiment with the close assistance assistance from Z. C., and X. P. Y., and in consultation with D. S. S. and M. Z. H.; SLS beamline support was provided by A. C. and V. N. S. Y. L., X. X. and S. J. synthesized and characterized the CoSi samples, H. B. and S. F. synthesized and characterized the RhSi samples. G. C. performed the first-principles/density functional theory calculations; D. S. S., T. A. C. and G. C. performed the analysis, interpretation and figure development in consultation with I. B., J.-X. Y, X. P. Y., Z. C., and M. Z. H.; D. S. S. and T. A. C. wrote the manuscript in consultation with G. C., I. B., J.-X. Y., and M. Z. H.; All authors contributed to revising and editing the manuscript. M. Z. H. supervised the project.

\clearpage
\clearpage

\clearpage

\clearpage

\begin{figure*}
\centering
\includegraphics[width=165mm]{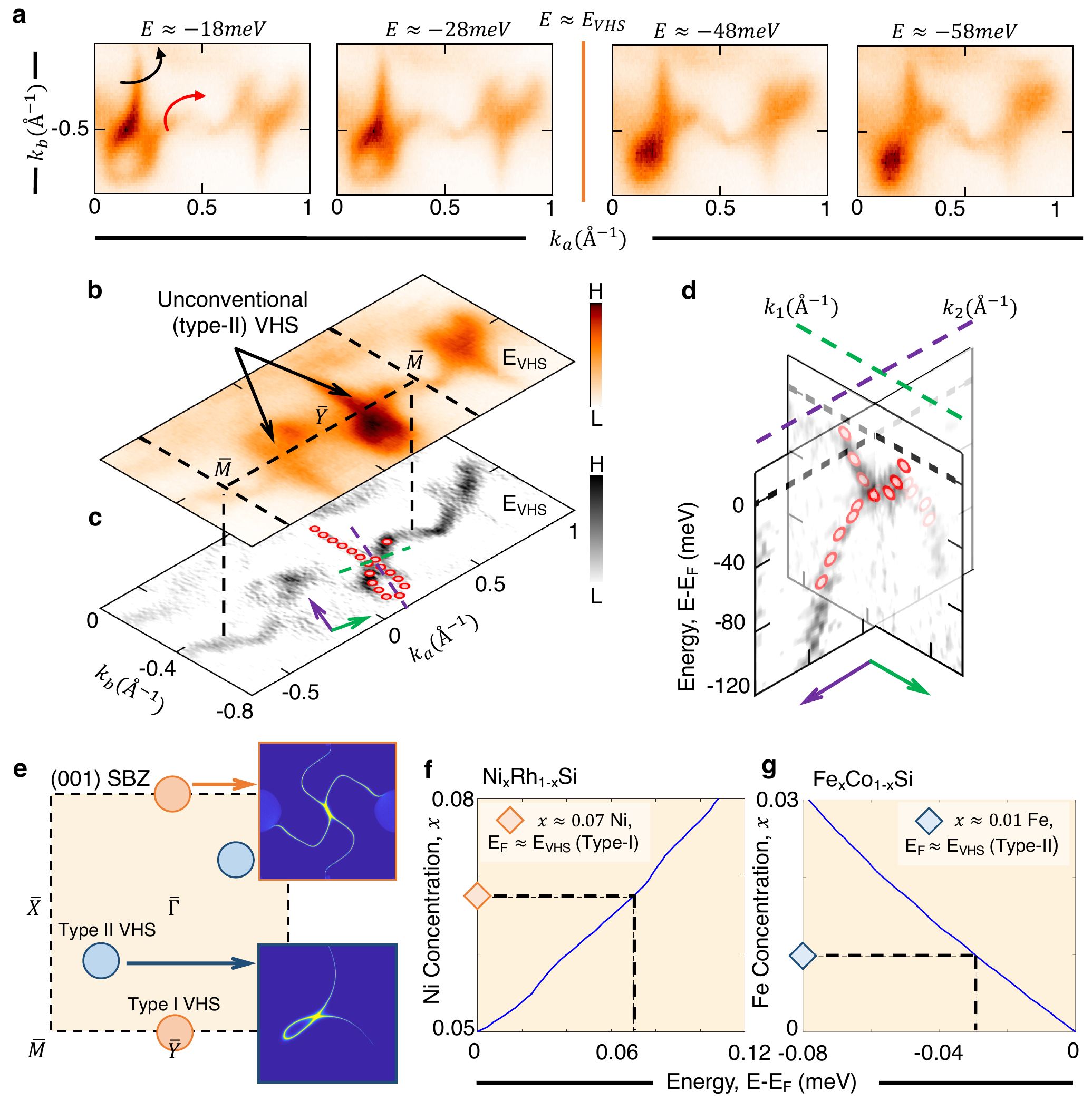}
\caption{\textbf{Tunable type-I and type-II helicoid-arc van Hove singularities in topological chiral conductor RhSi and CoSi.}
\textbf{a}, High-resolution VUV-ARPES measured constant energy contours near the $\bar{M}$ point for binding energies above and below the van Hove energy ($E=E_{VHS}$). The counterclockwise and clockwise white and red arrows, respectively, further illustrates the helicoid-winding property of the topological surface state near their termination points. \textbf{b}, Constant energy contour at binding energy $E \approx E_{VHS}$ (top panel) with a corresponding 2D curvature plot (\textbf{c}, bottom panel). A Lorentzian fitting procedure was used to track the helicoid-arcs trajectory across the surface BZ (red circles).}
\label{Fig4}
\end{figure*}

\addtocounter{figure}{-1}
\begin{figure*}[t!]
\caption{\textbf{d}, 2D curvature plot of the energy-dispersion cuts along perpendicular ($k_{a}, k_{b}$) directions intersecting the van Hove singularity in the surface Brillouin zone, as shown by the purple and green dashed-lines in panel (c). Red circles indicate the result of the Lorentzian fitting procedure used to track the band dispersion in the raw ARPES spectra. The observed saddle point dispersion provides direct spectroscopic evidence of an unconventional type-II van Hove singularity generated by the helicoid-arcs in CoSi. \textbf{e}, Cartoon illustration of the observable type-I and type-II van Hove singularities in RhSi (orange circles) and CoSi (blue circles), respectively. The inset further shows the the \textit{ab initio} calculated type-I and type-II van Hove singularities. \textbf{f}, Calculation of doping concentration as a function of binding energy for RhSi and CoSi in the chemical alloy Ni$_x$Rh$_{1-x}$Si and Fe$_x$Co$_{1-x}$Si, respectively. }
\label{Fig4}
\end{figure*}
\clearpage

\end{document}